%% file: EDM_Article_Submission.tex
\documentclass{edm_article}
\usepackage{booktabs}
\usepackage{url}
\usepackage{balance} 
\usepackage{hyperref}
\usepackage{tabularx}

\usepackage{xcolor}
\usepackage{soul} 

\makeatletter
\let\par\@@par
\makeatother

\begin{document}

\title{A Causal Framework for Estimating Heterogeneous Effects of On-Demand Tutoring}



%
%
%
%

\numberofauthors{6} 
%
\author{
\alignauthor Kirk Vanacore\\
    \affaddr{Cornell University}\\
    \email{kpv27@cornell.edu} 
            \and
\alignauthor Danielle R Thomas\\
    \affaddr{Carnegie Mellon University}\\ 
    \email{dchine@andrew.cmu.edu}
            \and
\alignauthor Digory Smith\\
    \affaddr{Eedi Labs}\\
    \email{digory.smith@eedi.com}
        \and
\alignauthor Bibi Groot\\
    \affaddr{Eedi Labs}\\
    \email{bibi.groot@eedi.com}
                \and
\alignauthor Justin Reich\\
    \affaddr{Massachusetts Institute of Technology}\\
    \email{jreich@mit.edu}
                \and
\alignauthor René Kizilcec\\
    \affaddr{Cornell University}\\
    \email{kizilcec@cornell.edu}
}

\maketitle

\begin{abstract}
This paper introduces a scalable causal inference framework for estimating the immediate, session-level effects of on-demand human tutoring embedded within adaptive learning systems. Because students seek assistance at moments of difficulty, conventional evaluation is confounded by self-selection and time-varying knowledge states. We address these challenges by integrating principled analytic sample construction with Deep Knowledge Tracing (DKT) to estimate latent mastery, followed by doubly robust estimation using Causal Forests. Applying this framework to over 5,000 middle-school mathematics tutoring sessions, we find that requesting human tutoring increases next-problem correctness by approximately 4 percentage points and accuracy on the subsequent skill encountered by approximately 3 percentage points, suggesting that the effects of tutoring have proximal transfer across knowledge components. These effects are robust to various forms of model specification and potential unmeasured confounders. Notably, these effects exhibit significant heterogeneity across sessions and students, with session-level effect estimates ranging from $-20.58pp$ to $+19.95pp$. Our follow-up analyses suggest that typical behavioral indicators, such as student talk time, do not consistently correlate with high-impact sessions. Furthermore, treatment effects are larger for students with lower prior mastery, which was true for both higher and lower socioeconomic status students.  This framework offers a rigorous, practical template for the evaluation and continuous improvement of on-demand human tutoring, with direct applications to other forms of on-demand assistance, including emerging AI conversational tutoring interventions.
\end{abstract}

\keywords{Tutoring, Adaptive Learning Systems, Causal Inference.}
\vspace{-0.5\baselineskip}

\section{Introduction}

Tutoring is widely recognized as one of the most effective educational interventions, particularly for students struggling academically \cite{dietrichson2017,guryan2023,nickow2024,kraft2024}. However, recent large-scale evidence suggests that the magnitude of K-12 tutoring effects varies widely across contexts, populations, and implementations, complicating efforts to generalize findings and guide policy decisions \cite{kraft2024Scaling}. Much prior work has focused on “high-dosage” tutoring, which is typically defined as sustained, scheduled sessions delivered multiple times per week \cite{robinson2021}. However, a growing class of interventions provides on-demand tutoring: brief support intended to deliver instruction precisely at moments of student difficulty \cite{deacon2023,wang2025}. These interventions are increasingly embedded within adaptive learning platforms and are delivered by human tutors or AI-based agents \cite{Prihar2022,wang2025}. 

These embedded tutoring systems generate rich, fine-grained interaction data, creating unprecedented opportunities for continuous evaluation and improvement. However, they also introduce a fundamental methodological problem: students self-select into tutoring at moments of struggle, creating severe time-varying confounding that undermines naive observational comparisons. As a result, despite the growth of embedded tutoring systems, the field lacks scalable methods for rigorously estimating the causal impact of individual tutoring interactions in real-world learning platforms.

This paper addresses this gap by introducing a scalable causal inference framework for estimating the effect of on-demand assistance -- in this case, human tutoring -- within an adaptive mathematics learning platform. Rather than evaluating entire programs or long-term interventions, the framework is designed to estimate the immediate causal effect of individual tutoring sessions on subsequent student performance. Our proposed framework integrates three components into a unified pipeline: (1) principled construction of analytic samples to approximate counterfactual comparisons, (2) latent knowledge state estimation using Deep Knowledge Tracing (DKT; \cite{piech2015dkt}) to address time-varying confounding, and (3) estimation of conditional average treatment effects for local, individual tutoring sessions using Generalized Random Forests (Causal Forests; \cite{athey2019grf}).
In combination, these components provide a practical template for estimating average treatment effects and effect heterogeneity from large-scale learning interaction logs.

We further demonstrate various robustness tests for this framework, showing that with and without contextual variables from outside of the platform (e.g., standardized assessments and demographic variables), this pipeline produced consistent estimates of the tutoring effects. Our framework addresses central challenges in observational studies of educational interventions, including self-selection into treatment, unmeasured baseline ability that changes as students learn, and treatment effect heterogeneity. Furthermore, it provides opportunities to better understand when, for whom, and why tutoring is effective. Finally, this method could be used to estimate outcomes for evaluating AI models for tutoring.

Applying our framework to data from an embedded tutoring system, we find that requesting tutoring has a positive impact on immediate performance by increasing the likelihood that students answer the next problem correctly by an average of $4.01$ percentage points ($pp$). We further find that this affects near transfer across skills: tutoring increases the probability of answering the first problem on the subsequent skill correctly by $2.73pp$. However, there is substantial variability in these effects, with local effect estimates ranging from  $-20.58pp$ to $19.95 pp$ for immediate performance correctness and from $-23.25 pp$ to $+18.24 pp$ for near transfer. We find that these estimates are robust to potential confounding from sample selection or unmeasured confounding. However, our marginally nonsignificant placebo suggests that these effect estimates may be biased towards underestimating the effect sizes. Students with lower estimated prior knowledge tended to benefit more from tutoring, although this relationship was slightly attenuated for lower-income students. We conclude by discussing how causal approaches of this kind can help researchers and practitioners better understand and evaluate the mechanisms that drive effective on-demand tutoring, ultimately informing the design of improved human and AI tutoring systems.
\vspace{-0.5\baselineskip}
\section{Background}
\vspace{-0.5\baselineskip}
\subsection{Efficacy and Heterogeneity in Tutoring Interventions}
Tutoring is widely recognized as one of the most effective academic interventions for improving student learning outcomes \cite{dietrichson2017,guryan2023,nickow2024}. Comprehensive meta-analyses have found that tutoring consistently yields significant positive effects on student learning \cite{kraft2024,nickow2024}. These effects are impactful across grade levels and subject areas, often outperforming other educational interventions such as class-size reduction or extended school days \cite{guryan2023,nickow2024}. However, the efficacy of tutoring is rarely uniform; effects are often heterogeneous across different implementations \cite{nickow2024}. Impact also varies significantly according to characteristics such as prior achievement, socioeconomic status, and baseline achievement \cite{dietrichson2017}. Although tutoring often targets lower-performing students to close achievement gaps \cite{guryan2023}, other studies on voluntary, on-demand models suggest that students with higher baseline engagement or fewer structural barriers are sometimes more likely to access and benefit from support \cite{aleven2016help,deacon2023}. Understanding this heterogeneity is crucial to determine whether specific tutoring interventions narrow achievement gaps or inadvertently widen them due to differential uptake.

The magnitude of tutoring effects is also inextricably linked to the implementation model. Much of the literature advocates for "high-dosage" tutoring, typically defined as sustained, scheduled sessions occurring three or more times per week, as a primary driver of efficacy \cite{kraft2024,nickow2024}. However, scaling such intensive models presents logistical and financial challenges. Consequently, many Adaptive Learning Systems (ALS) have adopted "on-demand" or "just-in-time" models, where brief support is triggered by students making specific errors rather than a fixed schedule \cite{Prihar2022,thomas2024,wang2025}. Although scalable, these embedded interactions complicate the traditional definition of dosage, as exposure to treatment becomes a function of student agency and immediate need rather than administrative assignment.

\subsection{Integrating Human Tutoring in Adaptive Learning Systems}

To address the scalability and cost constraints of high-dosage human tutoring, recent approaches have integrated human support directly into ALS. “Hybrid" or "human-in-the-loop" models aim to combine the personalized, immediate feedback of AI-driven software with the motivational and complex pedagogical support of human tutors \cite{chine2022,thomas2024}. Research suggests that human tutoring can enhance the benefits of ALS \cite{gurung2025}. For instance, human tutors—guided by real-time dashboards to intervene when students struggle—yields statistically significant additional boosts in time-on-task and skill proficiency within an ALS, compared to students working in an ALS alone \cite{gurung2025}. These hybrid approaches seem to benefit lower-performing students more than higher-performing ones, suggesting that human intervention may most benefit students who most need it \cite{thomas2024}.

Integrated tutoring models often take the form of "on-demand" chat support embedded within the learning environment. In these settings, the ALS identifies a knowledge gap or a specific misconception, and a human tutor enters the loop to provide targeted scaffolding \cite{harrison2023}. Evaluations of such systems have shown that students who engage with this hybrid support demonstrate higher learning gains and knowledge transfer than those using the ALS in isolation \cite{harrison2023,wang2025}. Furthermore, interventions explicitly designed to combine human tutoring with ALS personalization have been shown to nearly double math learning gains compared to control groups, providing a scalable mechanism to promote educational equity for marginalized students \cite{chine2022}. By offloading rote instruction to the ALS and reserving human capital for high-value interactions, these systems offer a promising pathway to deliver effective tutoring at scale \cite{chine2022,thomas2024,wang2024tutor}.

\subsection{Methodological Challenges in Estimating On-Demand Effects}

Although the efficacy of tutoring is well-documented, estimating the causal impact of on-demand help within adaptive systems presents unique methodological hurdles. Unlike scheduled "high-dosage" tutoring, where attendance is often mandated or fixed, "on-demand" support is driven by student agency. This introduces a specific form of confounding: students are more likely to ask for help precisely when they are least likely to succeed without it \cite{aleven2016help,robinson2021}. Consequently, naive comparisons between tutored and non-tutored problem attempts often yield negative correlations, reflecting the student's struggle rather than the intervention's failure.

To address this, prior EDM research has largely relied on two approaches: experiments/randomized controlled trials (RCTs) and observational matching strategies. While RCTs remain the gold standard, they are often implemented at the school or student level to evaluate general program access rather than specific interactions \cite{deacon2023,robinson2025effects,wang2025}. Randomizing individual help requests, which may require denying help to a struggling student for the sake of experimental control, is often ethically prohibitive and disruptive to the learning experience (e.g., wait-list designs where students are randomly assigned to get help right away or later are still disruptive). As a result, granular, problem-level insights must often be derived from observational data. In observational settings, researchers have traditionally employed propensity score matching to balance treatment and control groups on observed covariates such as prior achievement, demographics, and topic difficulty \cite{chine2022}. However, standard matching approaches typically rely on static or slowly changing variables (e.g., pre-test scores, diagnostic assessments). These measures fail to capture the highly dynamic, time-varying nature of a student’s knowledge state during a learning session. A student's decision to ask for help is often a function of immediate confusion or a specific misconception, which static or slowly-changing covariates cannot detect.

More recent work has attempted to incorporate dynamic measures of mastery into causal estimation. Approaches such as the "Rebar" and "ReLOOP" method have demonstrated that using predictions from high-dimensional models as covariates can reinforce matching estimators and reduce bias \cite{sales2018,pei2024boosting}. These methods allow for more precise effect estimation in the contexts of RCTs but are underutilized in observational studies where conditions are not randomized.

A related significant yet underdeveloped area in Educational Data Mining (EDM) is the integration of knowledge tracing (KT) into observational causal inference frameworks to account for time-varying confounding rooted in a student's historical performance \cite{abdelrahman2023,pardos2010modeling,pelanek2017bayesian}. In particular, DKT extends this by utilizing recurrent neural networks to generate high-dimensional latent states that capture complex temporal dependencies in student learning \cite{piech2015dkt}. However, even with these richer representations, standard regression or linear adjustments may not fully account for the complex, non-linear heterogeneity in how different students respond to tutoring. 

Finally, most causal studies in EDM and related fields focus on average treatment effects \cite{sales2018,pei2024boosting} or subgroup analyses \cite{kizilcec2017towards,kizilcec2020welcome,Pham_2024,Prihar2022}. A growing body of work also examines heterogeneity by how students interact with learning environments \cite{sales2020effect,pei2025fully,Vanacore_Sales_Liu_Ottmar_2023,vanacore2024effect}. However, the field would benefit from methods that estimate local, session-level estimates of effectiveness. These estimates could be leveraged to flexibly explore the circumstances under which ALS elements, like on-demand tutoring, are most effective, including elements of heterogeneity based on how students respond.


\vspace{-0.5\baselineskip}

\section{Current Study}
The current study proposes an observational approach for evaluating tutoring sessions embedded within digital learning platforms by estimating their impact on students’ subsequent performance. The approach integrates DKT–based estimates of students’ latent knowledge states into a causal inference framework, based on the assumption of strong ignorability, that combines causal forests for counterfactual outcome estimation with augmented inverse probability weighting to obtain doubly robust, session-level treatment effects. Together, these methods yield estimates of individual tutoring session effectiveness and enable downstream analyses that characterize heterogeneity in tutoring impacts across students, contexts, and instructional conditions. We use this framework on data from Eedi, an ALS that focuses on helping students overcome math misconceptions, during which students can access on-demand human tutoring through brief text chats.
\vspace{-0.5\baselineskip}
\section{Context}
\subsection{Data}
The data used for this analysis came from the treatment arm of a multi-year randomized controlled trial (RCT) designed to estimate the causal impact of Eedi on middle school mathematics achievement. Schools were randomly assigned to treatment or control conditions at the school level, with the treatment group receiving access to Eedi for two full academic years while control schools continued with business-as-usual instruction. Our current sample only includes the treatment arm of this RCT: the 12 schools and 2,585 students who had access to and used Eedi during the two-year study. 

The original RCT was conducted by an independent professional evaluation organization that was not involved in the present study. The evaluation organization followed an ethics protocol that included school consent, safeguards to ensure that study procedures did not deviate from typical daily instruction, and legally compliant data handling. All data transferred from Eedi to the researchers were fully deidentified. The current study was also approved by Cornell's institutional review board (IRB0149250).

\subsection{Eedi} This study was run in Eedi, which is a digital mathematics platform designed to support middle school students by identifying and addressing misconceptions in real time through diagnostic assessment. Eedi uses a large bank of carefully engineered multiple-choice questions, where each incorrect option maps to a specific, well-documented mathematical misconception. This design allows Eedi to pinpoint gaps in understanding at a fine grain and respond with targeted support, including short video explanations, structured practice, and follow-up diagnostic questions. Teachers can deploy the system flexibly in lessons or as homework, using its analytics dashboards to see patterns of misunderstanding across individuals and classes and to adapt instruction accordingly. Students work independently on quizzes of five questions assigned by their teachers. Based on their answers, Eedi responds adaptively with a sequence of hints, videos, and fluency practice to help students overcome misconceptions. Evidence from a large, school-level randomized controlled trial suggests that the platform can produce modest but meaningful improvements in mathematics attainment when implemented over an academic year \cite{harrison2023}.

\subsection{On-Demand Tutoring in Eedi.} Within Eedi, human tutoring was implemented as on-demand, synchronous, one-to-one chat-based support integrated directly into students’ ongoing problem-solving activity. At any point, students could initiate a live session that immediately connected them with an expert tutor. Tutors received structured context at the start of each session—including the problem text, the student’s answer, and the associated misconception, allowing them to provide targeted, dialog-based support focused on diagnosing and resolving misunderstandings before students returned to independent work.

The intervention involved seventeen tutors, each with at least three years of teaching experience and specific training in tutoring. Sessions were designed to be brief and task-focused, enabling students to move fluidly between independent practice and individualized help. Tutors communicated exclusively through text chat and used a Socratic approach that emphasized questioning and guided reasoning rather than direct answer-giving. Sessions concluded once students demonstrated sufficient understanding or chose to resume their lesson, positioning tutoring as a scalable, just-in-time complement to the broader Eedi's learning environment.

Table~\ref{tab:tutoring_desc} summarizes the characteristics of the tutoring sessions. Tutoring sessions were relatively brief, with a median duration of 4.2 minutes and 14 messages exchanged. Tutors contributed approximately 60\% of messages on average, consistent with their role in guiding the dialogue.

\begin{table}[ht]
\centering
\footnotesize
\caption{Tutoring Session Characteristics}
\label{tab:tutoring_desc}
\begin{tabular}{lcccc}
\toprule
\textbf{Metric} & \textbf{Mean} & \textbf{Median} & \textbf{Q25} & \textbf{Q75} \\
\midrule
Total Messages & 17.7 & 14 & 9 & 23 \\
Tutor Messages & 10.5 & 8 & 5 & 14 \\
Student Messages & 7.2 & 6 & 3 & 9 \\
Duration (minutes) & 5.4 & 4.2 & 2.4 & 6.9 \\
\bottomrule
\end{tabular}
\end{table}

\begin{figure*}[ht]
\centering
\includegraphics[width=.8\textwidth]{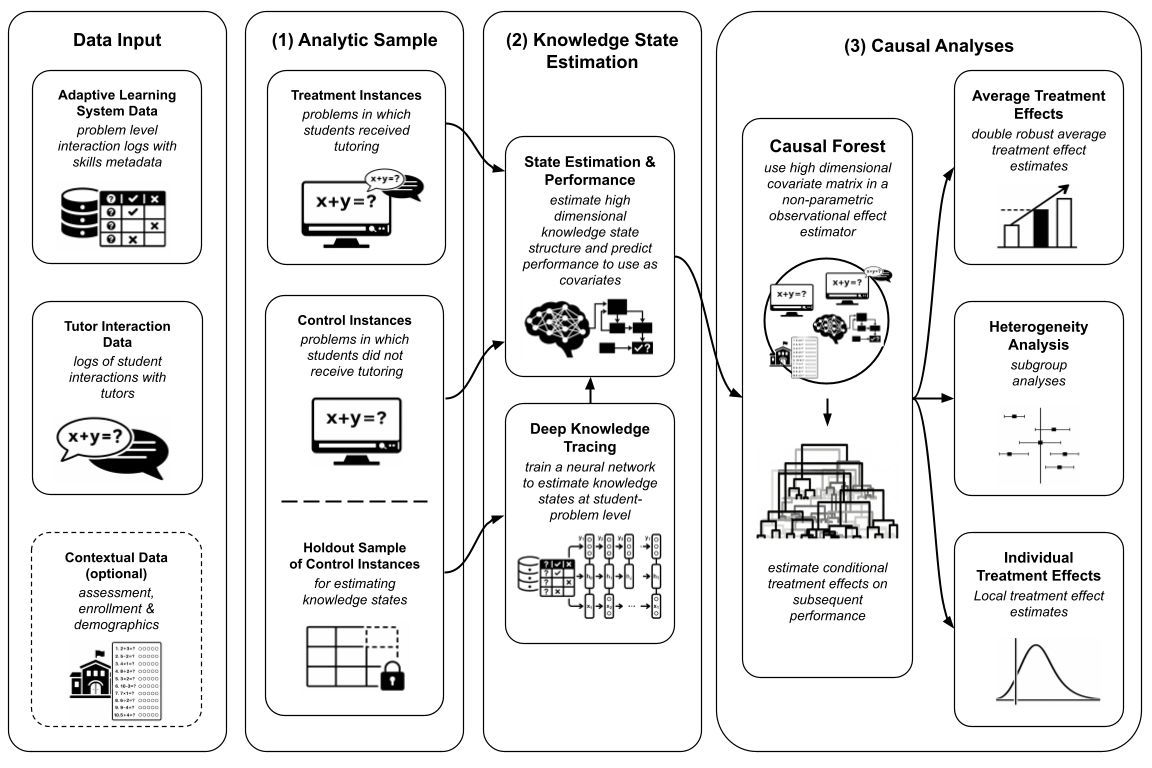}
\Description{A schematic overview of a causal framework for estimating heterogeneous effects of on-demand tutoring. The figure shows multiple data sources, including student learning-system logs, tutoring interactions, contextual variables, and assessment outcomes, feeding into an integrated analytic pipeline. The pipeline estimates student knowledge states, models causal effects of tutoring, and produces downstream analyses of average effects, heterogeneous treatment effects, robustness checks, and explanatory factors associated with variation in tutoring impact.}
\caption{\textbf{Causal Framework for Estimating Heterogeneous Effects
of On-Demand Tutoring.} The pipeline integrates ALS and tutor interaction logs to estimate student knowledge states and causal effects. These estimates drive downstream insights into effect variability. While the framework functions independently of contextual data, we incorporate it here to test robustness and identify variables for heterogeneity analysis.}
\label{fig:TutorImpact}
\end{figure*}
\vspace{-0.5\baselineskip}

\section{Methods: Causal Framework for Local Effect Estimation}
Our causal inference framework consists of three interconnected stages, illustrated in Figure \ref{fig:TutorImpact}. (1) First, we construct an analytic dataset that links each tutoring intervention to its immediate outcome while defining an appropriate control group. (2) Second, we train a DKT model on a held-out sample of students to generate latent representations of student knowledge for both the treatment and control samples. (3) Third, we apply a Causal Forest estimator that jointly estimates propensity scores and treatment effects using the DKT-derived hidden states and predicted probabilities as covariates. We validate our findings through a series of robustness checks that incorporate contextual covariates (e.g., school assignment, standardized assessments, and demographics), alternative sample configurations to address potential participation bias, and a placebo outcome test. This framework addresses three central challenges in observational tutoring research: confounding from baseline knowledge differences, self-selection into treatment, and heterogeneity in treatment effects across students.

\subsection{Analytic Sample Construction}
\label{sec:sample}
Our method requires three samples: treatment, control, and hold-out. Table \ref{tab:sample_flow} presents the sample breakdown. When a student received tutoring on a problem, that instance entered the treatment sample. The next problem the student attempted without help served as our \textit{immediate performance} outcome, and the next problem on a new skill served as our \textit{near transfer} outcome. To ensure comparability, only problems for which at least one student received tutoring were included in the causal analysis. In addition, problem attempts from treated students in which they did not receive tutoring were excluded from the primary analysis. Treated observations were also excluded when no subsequent problem attempt was available to serve as an outcome.

To avoid potential Stable Unit Treatment Value Assumption (SUTVA) violations \cite{Rubin1974}, only students who never received tutoring were eligible for the control sample, as prior exposure to tutoring could influence later outcomes. We relax this restriction in a robustness check to assess whether this sample construction meaningfully affects results (see Section \ref{sec:robust}). Students who were never treated were randomly divided between the control and hold-out samples. The hold-out sample was used exclusively for DKT estimation (Section \ref{sec:DKT}), while the control sample was reserved for causal effect estimation. The analytic control sample, therefore, consisted of students who (a) never received tutoring and (b) attempted at least one of the same problems as students in the treatment sample. Because not all problems overlapped with tutoring interventions, control-group attempts on non-overlapping problems were excluded. Students who lacked any problem attempts overlapping with the treatment sample were removed from the final analytic control sample.

\input{06_tables/sample_flow_table}

\subsection{Knowledge State Estimation}
\label{sec:DKT}

We implemented a DKT model using a Long Short-Term Memory (LSTM) architecture to generate student knowledge estimates used as pre-treatment covariates in the causal analysis \cite{piech2015dkt}. The model uses a student’s sequential history of question attempts to predict the probability of correctly answering subsequent questions. Critically, the model maintains a latent hidden state that encodes information from prior performance across time, allowing it to summarize a student’s evolving knowledge state even when past interactions involve different constructs than the one being predicted.

The use of DKT serves a substantive causal purpose. Student knowledge is a key confounder in this setting: students are more likely to seek or be offered tutoring when they are struggling, and baseline knowledge strongly predicts future performance. By conditioning on a rich, multidimensional representation of students’ latent knowledge states, we aim to reduce bias arising from time-varying confounding in tutoring assignment and outcome generation. Furthermore, DKT can provide specific point-in-time estimates of baseline knowledge that account for recent learning, which may be more robust than global pretest measures.

We implemented the DKT model in PyTorch following the architecture first introduced by Piech et al. \cite{piech2015dkt}. The training data included 197,332 problem attempts across 7,485 problems representing 1,598 distinct knowledge components. Item embeddings were passed to an LSTM with a hidden dimension of 50, which updated its hidden state $h_t$ at each time step. The resulting 50-dimensional hidden state vector $(H_1, \ldots, H_{50})$ was used as the primary knowledge representation in our causal analysis. The output layer applied a sigmoid activation function to produce probability estimates for the next construct. The model achieved an acceptable AUC of 0.72 on the analytic control sample (which was not used for training), consistent with typical performance reported for DKT models \cite{abdelrahman2023}.

\subsubsection{Feature Extraction.} For each observation in the treatment and control samples, we extracted the 50-dimensional LSTM hidden state calculated \textit{just prior to the intervention.} We also included the predicted performance on the current problem, in which the student could receive tutoring support, and the subsequent problem, which serves as an outcome. Including these as covariates in causal analysis is modeled after the \textit{Rebar} method, which has been shown to produce precise, unbiased effect estimates by applying models generated from data outside of the treatment and control sample to those samples \cite{sales2018}.

\vspace{-0.5em}
\subsection{Causal Forest Estimation}
We estimated treatment effects using Generalized Random Forests (GRF) as implemented in the \texttt{grf} R package \cite{athey2019estimating}. We chose to incorporate causal forests into the pipeline because of their nonparametric flexibility and their ability to model high-dimensional treatment effect heterogeneity \cite{athey2019grf}. 

Furthermore, causal forests offer three other advantages that are particularly relevant in our setting. First, they have been shown to perform well under targeted selection, when treatment assignment depends on predictors of the outcome itself \cite{Hahn2020,athey2019grf}. In the context of on-demand tutoring, students are more likely to seek or receive tutoring precisely when they anticipate poor performance absent intervention, rendering treatment assignment endogenous and outcome-driven. Causal forests are explicitly designed to accommodate this form of selection. Second, a key feature is \emph{honesty}, in which separate subsamples are used for tree construction and treatment effect estimation. This sample-splitting reduces adaptive overfitting and supports valid inference by producing out-of-sample counterfactual predictions analogous to leave-one-out estimation \cite{Wu2018}.  Third, causal forests directly estimate Conditional Average Treatment Effects (CATEs) as flexible functions of a high-dimensional covariate space. Given the richness of our covariates—including the 50-dimensional DKT hidden state representation—this approach enables estimation of fine-grained, unit-specific conditional effects and supports downstream analyses of treatment effect heterogeneity across students and contexts.

\subsubsection{Model Specification.} The covariate matrix $X$ included 50 DKT hidden state dimensions, cumulative historical accuracy, and DKT-based probability estimates of correct responses. The forest was trained with 500 trees, honesty enabled, and hyperparameters selected via cross-validation. We specified students as clusters in the causal forest, implementing cluster-robust estimation such that all observations from a given student are assigned to the same subsample during tree construction. As a result, treatment effect estimates account for within-student dependence, and predictions for each student are generated without using that student’s own data.

\subsubsection{Estimation Engine: Robinson Decomposition and Propensity Modeling}

We estimate heterogeneous treatment effects using Causal Forests, which rely on the Robinson decomposition (also known as a residual-on-residual or orthogonalization strategy). Let $Y_i$ denote the observed outcome for unit $i$, $Z_i \in \{0,1\}$ the treatment indicator, and $X_i$ the vector of pre-treatment covariates. For each outcome definition (next-problem correctness and next-skill correctness), the procedure first fits two functions using separate regression forests: the conditional mean outcome model
\[
\hat{m}(X) = \mathbb{E}[Y \mid X]
\]
and the propensity score model
\[
\hat{e}(X) = \mathbb{P}(Z=1 \mid X).
\]

These estimates are then used to construct orthogonalized (residualized) variables:
\begin{equation}
    \tilde{Y}_i = Y_i - \hat{m}(X_i), 
    \qquad
    \tilde{Z}_i = Z_i - \hat{e}(X_i),
\end{equation}
where $\tilde{Y}_i$ represents the deviation of the observed outcome from its covariate-expected value, and $\tilde{Z}_i$ represents the deviation of the observed treatment assignment from its predicted probability. This transformation removes outcome variation explained by baseline covariates and treatment variation explained by selection on observables, yielding Neyman-orthogonal signals for effect estimation.

The Causal Forest is then trained to model the relationship between $\tilde{Y}_i$ and $\tilde{Z}_i$ as a function of $X_i$, which isolates the causal signal after adjusting for baseline outcome differences ($\hat{m}$) and treatment selection ($\hat{e}$).

Propensity scores are therefore estimated internally as part of the Causal Forest procedure via a regression forest for $\hat{e}(X)$. Valid estimation requires overlap (positivity), meaning each unit must have a non-negligible probability of both treatment and control conditional on $X$. In our analytic sample, estimated propensity scores ranged from 0.03 to 0.89, with most observations falling between 0.05 and 0.20. Appendix A.1 presents the full propensity score distribution and associated overlap and robustness diagnostics.

\begin{equation}
\tilde{Y}_i = \tau(X_i)\tilde{Z}_i + \epsilon_i
\end{equation}

\subsubsection{Effect Estimates.} We report three types of effects. The Conditional Average Treatment Effect (CATE) provides unit-level predictions, where $\tau(x)$ represents the difference in expected potential outcomes for an individual with covariates $x$. Specifically, it captures the contrast between the expected outcome if the student receives tutoring, $Y^{(1)}$, and the expected outcome if the student does not receive tutoring, $Y^{(0)}$:
\begin{equation}
    \tau(x) = E[Y^{(1)} - Y^{(0)} | X = x]
\end{equation}

The Average Treatment Effect (ATE) is the population-level effect, computed using Augmented Inverse Probability Weighting (AIPW) scores:
\begin{equation}
    \hat{\tau}^{\mathrm{ATE}}
=
\frac{1}{n}\sum_{i=1}^n
\left[
\hat{\tau}(X_i)
+
\frac{Z_i - \hat{e}(X_i)}{\hat{e}(X_i)(1-\hat{e}(X_i))}
\bigl(Y_i - \hat{m}_{Z_i}(X_i)\bigr)
\right]
\end{equation}
The Average Treatment Effect on the Treated (ATT) is determined by restricting this calculation to treated units and represents the average effects for those who chose the treatment. The use of AIPW makes the ATE and ATT 'doubly robust,' as they combine the outcome predictions with propensity weighting to correct for bias \cite{Glynn2010}. Thus, if either the outcome model or the propensity score model is misspecified, the estimates remain consistent and asymptotically unbiased.

\subsubsection{Heterogeneity Analysis.} \label{sec:hetero}To test for moderators of the treatment effect, we treat the estimated unit-level CATEs ($\hat{\tau}_i$) as the outcome variable in a Linear Mixed Model (MLM):
\begin{equation}
    \hat{\tau}_i = \beta_0 + \beta_1 \cdot \text{Moderator} + \mu_s + \epsilon_i
\end{equation}
where $\mu_s$ accounts for student-level clustering. This allows us to determine (via significance testing) which moderators explain the variability in treatment effects. Notably, we do not rely on more standard causal-forest heterogeneity summaries such as best linear projection, because our substantive interest extends to moderators that are not included in the causal forest itself, including variables derived from the tutoring sessions. Instead, we use the estimated CATEs as outcomes in follow-up analyses to examine how these external variables relate to treatment effect variation.
\input{06_tables/ate_results_table}

\subsection{Robustness and Sensitivity Checks}
\label{sec:robust}
To assess whether the model is robust to misspecification and potential unobserved confounders, we conducted a series of robustness checks. Our first sets of robustness checks determine whether adding different variables or sample configurations influences the results. 

\subsubsection{External Variables Added} First, we incorporated additional covariates not derived from platform usage. These included standardized pretest scores of mathematical knowledge (NWEA MAP RIT), which provide an external measure of baseline ability independent of platform performance; a gender indicator with three categories (male, female, or neither); and Pupil Premium Funding (PPF) eligibility, a government designation for students from disadvantaged backgrounds that serves as a proxy for socioeconomic status. We also included school identifiers. The goal of this analysis was to test whether the estimated Average Treatment Effect (ATE) remains stable after adding external covariates, which would suggest that the DKT-based features may adequately capture relevant confounding and that the estimated treatment effect is not driven by unobserved demographic or institutional factors. 

\subsubsection{Washout Sample Included} The second robustness test evaluates potential selection bias arising from defining the control group as students who have \textit{never} requested tutoring. These students may differ from treated students in ways not fully captured by DKT features or other observed covariates. Although restricting the control group in this way helps prevent SUTVA violations (see Section \ref{sec:sample}), it may introduce confounding if never-treated students are fundamentally different from those who seek tutoring. To examine this possibility, we re-estimate the model using an expanded control group that includes students who previously received tutoring, provided that the prior session did not occur within the previous two skills (i.e., a washout sample). This specification allows us to account for potential carryover effects while reducing selection bias by increasing overlap between treatment and control observations.

\subsubsection{Pre-Intervention Placebo Test} Next, we conducted a placebo test using an outcome for which no causal effect is possible: student performance on a problem that occurs \textit{before} the tutoring intervention. Placebo or “negative control” outcomes are widely used in causal inference as a diagnostic tool for detecting residual confounding or model misspecification. If the identification strategy is valid, estimated treatment effects on such outcomes should be null, since they temporally precede treatment and cannot be causally influenced by it \cite{eggers2024placebo}. We chose a problem three items before the intervention as our placebo outcome to avoid any interference with the intervention. Finding no significant effect on this pre-treatment outcome would therefore provide additional evidence that the method is not simply capturing baseline differences between students who choose tutoring and those who do not, but is instead identifying a genuine causal effect of tutoring.

\subsubsection{Additional Checks}
Finally, we employ two additional robustness checks that are common in quasi-experimental analyses. We present these in detail in Appendix A. We tested the sensitivity of our results to propensity score model specification and trimming of the distribution tails to ensure common support \cite{crump2009dealing}. To quantify the robustness of our estimates to unmeasured confounding (i.e., omitted variable bias), we calculate the Robustness Value (RV). The RV represents the minimum strength of association that an unobserved confounder would need to have with both the treatment and the outcome (measured in terms of partial $R^2$) to reduce the estimated treatment effect to zero. This allows us to benchmark the required strength of an unmeasured confounder against observed covariates like prior mastery or student SES. 


\vspace{-0.5\baselineskip}
\section{Results}

\subsection{Treatment Effect Estimates}

Table~\ref{tab:ate_results} presents the doubly robust average treatment effect estimates from the Causal Forest analysis, with p-values adjusted for multiple comparisons. We find a statistically significant positive effect of tutoring on immediate performance and near transfer. The estimated average treatment effect (ATE) of tutoring on next problem correctness was $4.01$ percentage points ($pp$) (CI = [2.51, 5.51]), $p < 0.001$) and accuracy on the first attempt at a problem on the next skill was $2.73pp$ (CI = [1.12, 4.35]). The effects on the treatment group (ATT) were also significant and similar in magnitude.

\begin{figure*}[ht]
\centering
\includegraphics[width=.9\textwidth]{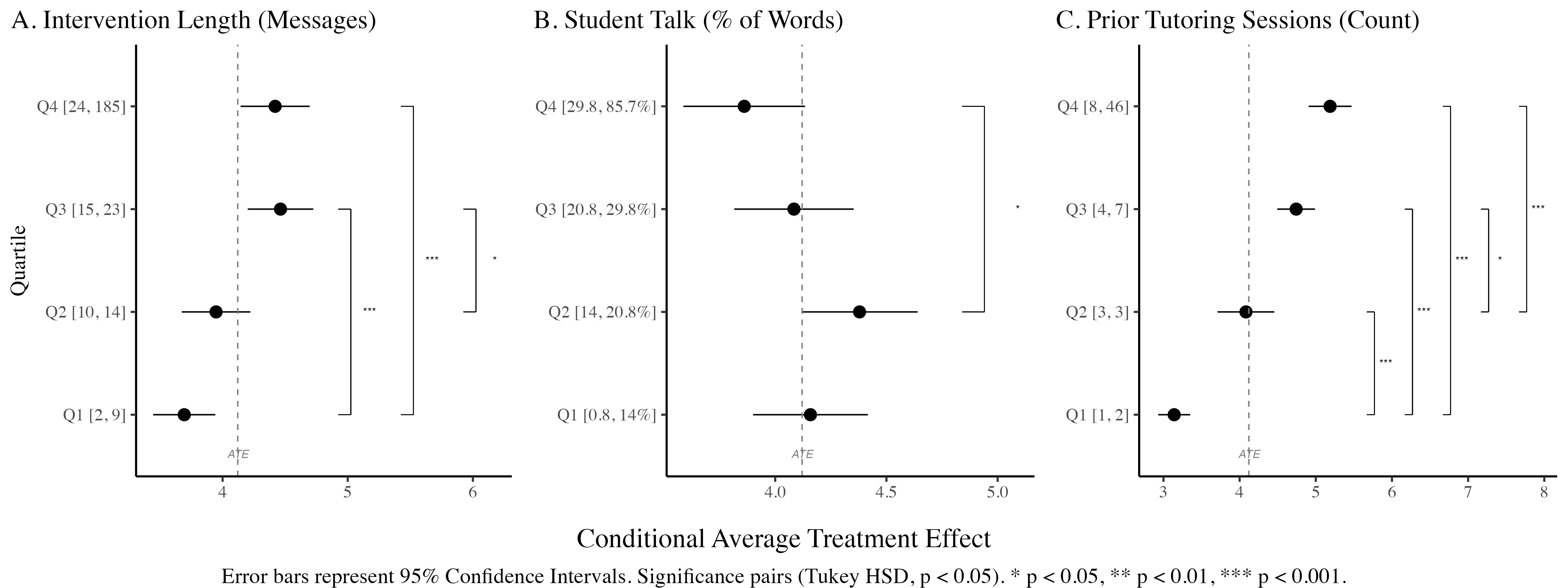}
\Description{Three-panel forest plot showing conditional average treatment effects on immediate performance by tutoring session characteristics. Panel A shows effects by intervention length in messages, with quartiles Q1 through Q4. Effects are lowest for the shortest sessions and higher for longer sessions, with confidence intervals shown as horizontal bars and significance brackets marking pairwise differences. Panel B shows effects by student talk as a percentage of words, with quartile-specific estimates and confidence intervals clustered around the overall average treatment effect. Panel C shows effects by the number of prior tutoring sessions, with effects increasing across prior-session quartiles, especially for students with more previous tutoring experience. Each panel includes a vertical dashed line marking the average treatment effect and horizontal error bars representing 95 percent confidence intervals.}
\caption{Effect Sizes on Immediate Performance by Tutoring Session Characteristics. Each effect is decomposed by quartiles of the characteristics. Quartile ranges are presented in brackets.}
\label{fig:Engagement_Panel}
\end{figure*}

\subsubsection{Robustness and Sensitivity.}

Table~\ref{tab:ate_results} presents results from several robustness checks designed to evaluate whether specific modeling choices, such as covariate selection and sample construction, influence the stability of the effect estimates. The inclusion of external demographic covariates yielded a nearly identical ATE of $3.93pp$ ($CI = [2.73, 5.51]$). This stability suggests that our Deep Knowledge Tracing (DKT)-derived features successfully capture the requisite student heterogeneity, rendering additional demographic data redundant for the purposes of confounding control. In the model specification that included the washout sample (the control included students who have been treated in the past, but not within two skills), students within the control group produced a lower but still statistically significant estimate of $2.17pp$ ($CI = [1.17, 6.16]$). The convergence and overlapping confidence intervals across these disparate design assumptions provide robust evidence for a consistent causal effect. These checks were replicated for the near-transfer outcome with similarly significant results.

The effect estimate for the placebo outcome—a measure theoretically unaffected by the treatment—was negative and non-significant; however, this is only after applying a Bonferroni correction to account for family-wise error. This result suggests that, if anything, students in the treatment group may actually be biased toward lower baseline performance consistent with targeted selection into treatment; consequently, the observed positive treatment effects are unlikely to be artifacts of model-driven upward bias.

Two additional robustness tests are detailed in Appendix A.1. We utilized paired matching as a diagnostic tool for propensity score estimation, finding that a Random Forest propensity model achieved superior covariate balance compared to linear specifications. Furthermore, we confirmed that trimming the tails of the propensity distribution to ensure stricter overlap did not significantly alter the treatment effect estimates.

\textbf{Unmeasured Confounding.} Finally, as discussed above, the framework depends on a strong ignorability assumption: after conditioning on observed covariates, there are no substantial unmeasured factors that jointly influence tutoring usage and outcomes. This matters here because plausible confounders, such as student motivation, are not directly observed. We assume that some of their influence is expressed through prior performance and interaction histories and is therefore partially captured by the DKT-based covariates, though residual unmeasured confounding may still remain. To test this assumption, we conducted a formal sensitivity analysis \cite{cinelli2020making} by calculating the Robustness Value ($RV_{q=1}$) required to fully explain away the estimated effect.\footnote{See Appendix A.2} Our results indicate that an unmeasured confounder would need to be at least three times as predictive of both treatment assignment and the outcome as the stroungest predictors (DKT outcome prediction, average accuracy, or cumulative attempts) to nullify the current findings. This level of confounding is unlikely in practice, suggesting the results are robust to plausible omitted variables.

\subsection{Treatment Effect Heterogeneity}

\begin{figure*}[ht]
\centering
\includegraphics[width=.65\textwidth]{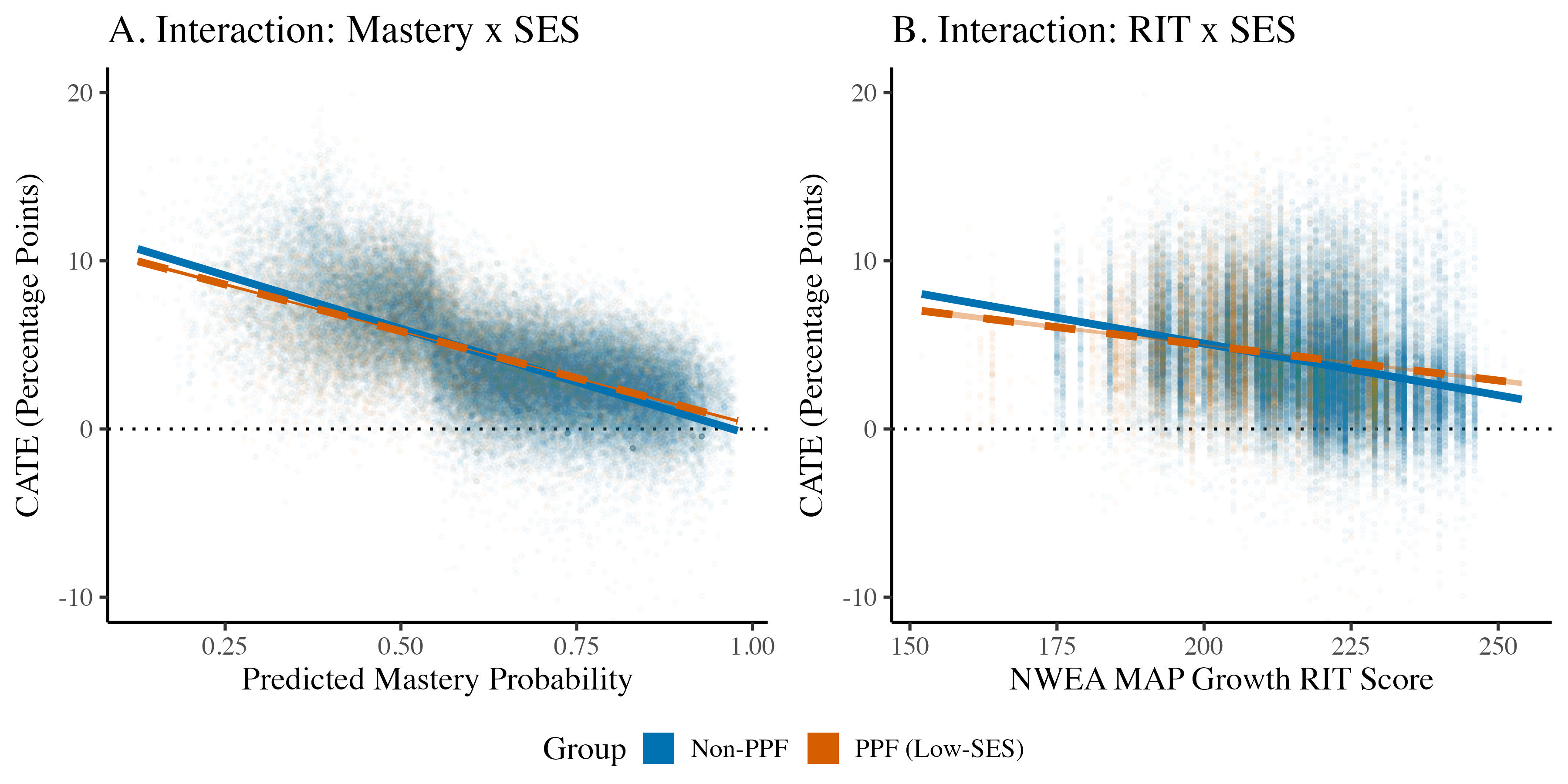}
\Description{Two-panel scatterplot showing heterogeneous treatment effect estimates by prior knowledge and socio-economic status. Panel A plots CATE in percentage points against predicted mastery probability, with separate trend lines for non-PPF students and PPF low-SES students. Both groups show declining estimated tutoring effects as predicted mastery increases. Panel B plots CATE in percentage points against NWEA MAP Growth RIT score, again with separate trend lines for non-PPF and PPF students. Both groups show declining estimated tutoring effects as RIT score increases, with the PPF trend slightly above the non-PPF trend at higher scores. A horizontal dotted line marks zero effect in both panels.}
\caption{Interactions between prior knowledge in terms of predicted mastery (A) or standardized test scores (B) and socio-economic status (SES) based on Pupil Premium Funding (PPF) eligibility.}
\label{fig:Heterogeneity_Panel}
\end{figure*}

While the average effects are positive, we observe substantial heterogeneity in local treatment effects across intervention sessions. The CATE for immediate performance has a standard deviation of $3.53pp$, and ranges from $-20.58pp$ to  $19.95pp$.  The CATE for near transfer has a standard deviation of $3.61pp$, and ranges from $-23.25pp$ to  $18.24pp$. Notably, there is a moderate correlation between these effects (\textit{r} = 0.46, \textit{p} < 0.001), suggesting that the immediate performance CATE may be a good, however imperfect, surrogate measure for near transfer; if a tutoring session has a positive effect on next problem correctness, it is likely to have a positive effect on the students' ability to solve problems on other similar skills as well.

\subsubsection{Heterogeneity by Tutoring Session Characteristics}

To illustrate how this method can be used to examine why some tutoring sessions are more effective than others, we analyzed three potential moderators: session length, proportion of student talk, and the number of prior tutoring sessions. We chose these moderators because each of them has been studied in more traditional tutoring environments. Longer and more frequent sessions are consistently pointed to as hallmarks of high-impact tutoring \cite{robinson2021}. Eliciting student talk is often associated with higher quality tutoring sessions \cite{chi2001learning}.  Notably, we view this as a preliminary analysis into understanding what might drive effectiveness in on-demand tutoring, which will lead to future directions for this line of work. 

Because these variables were non-normally distributed, each was divided into quartiles and compared using the heterogeneity model (Section \ref{sec:hetero}) with Tukey HSD post-hoc tests.\footnote{We also ran the models with transformed variables and found similar results, but chose to present the results in terms of quartiles to ease interpretation.}  Figure \ref{fig:Engagement_Panel} presents the differences in effect size on immediate performance by quartile (quartile descriptive statistics and results tables are presented in Appendix C).

Session length was measured by the number of messages sent between the tutors and students that showed the clearest and most consistent pattern. Moderately long sessions (Q3) produced significantly larger effects than shorter sessions (Q1 and Q2; $p < 0.001$), indicating that brief interactions are often insufficient to support learning. Sessions in the longest quartile (Q4) also outperformed the shortest sessions (Q1, $p < 0.05$), but did not yield reliably greater benefits than moderately long sessions, suggesting diminishing returns beyond a certain point. 

As a proxy for student talk, we used the percentage of total words in the session that were sent by the student. This measure of the proportion of student talk exhibited minimal and inconsistent associations with tutoring effectiveness. Only one comparison reached significance, indicating slightly lower effects in sessions with the highest levels of student talk (Q4 vs. Q2; $p < 0.05$), with all other contrasts non-significant. Students who sent between $14\%$ and $20\%$ of the words during the session had higher estimated effects than those who sent between $29.8\%$ and $85.7\%$. This contrasts with much of the work on student talk in tutoring, which often finds that getting students to engage is an important part of tutoring \cite{chi2001learning}. However, this analysis suggests that student talk may be less important in on-demand sessions that focus on specific problems. 

The number of tutoring sessions that preceded the intervention session was strongly and positively associated with impact. Almost all comparisons show that students with more previous sessions experienced substantially larger gains than first-time or infrequent users, with effect sizes increasing monotonically across quartiles. Taken together, these results suggest that tutoring is most effective when sessions are of at least moderate length and when students have prior experience engaging with on-demand support, whereas simply increasing the amount of student talk does not reliably translate into greater effects on proximal performance.

\subsubsection{Heterogeneity by Student Characteristics}

The CATE estimations also provide opportunities to evaluate which students benefit more from this form of tutoring. To assess this variability, we estimated a series of regression models predicting session-level CATEs (the full table of the model output can be found in Appendix B). Students with lower baseline ability exhibited larger treatment effects, with DKT mastery showing the strongest association ($\beta = -1.87$, $p < 0.001$), indicating that a one standard deviation decrease in predicted mastery corresponded to a 1.87 percentage point increase in effect size. Once DKT was included, NWEA RIT scores showed only a small positive association ($\beta = 0.08$), suggesting that the more proximal DKT-based measure of knowledge was the primary driver of heterogeneity. This pattern is consistent with both a ceiling effect—where higher-performing students have less room to improve—and the hypothesis that just-in-time tutoring is most valuable for students who are struggling. 

Overall, low-SES students benefited from tutoring, but to a slightly lesser extent than their peers ($\beta = -0.21$, $p < 0.05$) and interaction terms revealed that the relationship between prior knowledge and treatment effects was weaker for these students (DKT $\times$ PPF: $\beta = 0.14$, $p < 0.001$; RIT $\times$ PPF: $\beta = 0.33$, $p < 0.001$).  Figure \ref{fig:Heterogeneity_Panel} presents a visualization of this interaction. While lower-performing students generally benefited more from tutoring, this gradient was attenuated among low-SES students. However, as the figure shows, tutoring effects remain strongly related to prior knowledge: on average, lower-performing, low-SES students still appear to benefit more from tutoring than higher-performing, higher-SES students.

\vspace{-0.5\baselineskip}
\section{Discussion \& Conclusion}
This paper introduces a scalable causal framework for estimating problem-level effects of on-demand human tutoring embedded in an adaptive learning system. Across specifications, requesting tutoring produces modest but reliable gains in immediate performance and near transfer. 

One central contribution of this work is methodological. We show that time-varying knowledge representations derived from platform logs can be used as pretreatment covariates to address confounding driven by evolving mastery and task difficulty in observational studies of an opt-in feature. By integrating Deep Knowledge Tracing features \cite{piech2015dkt} into a doubly robust causal forest estimator \cite{wager2018,athey2019grf,Glynn2010}, the framework targets the core selection problem that tutoring is typically requested when students anticipate poor performance. The stability of ATE estimates after adding external covariates (test scores, SES proxy, school identifiers) suggests that platform-derived learning traces capture much of the relevant confounding structure. However, the marginally non-significant negative placebo effect may suggest some bias in the approach, but, in this application, that bias is likely an underestimation of the effect.

A key substantive result is that the ATE alone masks wide heterogeneity, including strongly positive and negative session-level effects. Although the average tutoring effect is comparable to lower-cost ALS features such as on-demand hints \cite{gurung2023common,Prihar_Syed_Ostrow_2022,Prihar_2021}, individual effects range from roughly $+20$ to $-20$ percentage points. This indicates that some tutoring sessions are highly effective while others are neutral or counterproductive.  

Our analysis of this heterogeneity provides a preliminary, though not comprehensive, attempt to identify the factors associated with higher-impact tutoring. Evaluating session characteristics highlights several potential drivers of effectiveness. Session length exhibits the clearest pattern: moderately long sessions (outperform very short sessions, while the longest sessions do not reliably improve upon moderately long ones. This suggests a diminishing marginal return on tutor time beyond a certain threshold. It is also possible that longer conversations indicated the need for additional remediation beyond what can be provided through a chat interface.

Furthermore, prior intervention experience is strongly and monotonically associated with larger impacts, indicating that students may 'learn how to use' tutoring more effectively over time. Alternatively, students who access tutoring more frequently may possess inherent characteristics, such as higher baseline engagement, that render the tutoring more beneficial. In contrast, the proportion of student talk shows minimal associations with impact. This finding is directionally inconsistent with classic tutoring accounts that emphasize student explanation and constructive engagement as drivers of learning \cite{chi2001learning}, and suggests that, in brief, problem-focused, on-demand chats, the quantity of student talk may be less informative than the specificity, timing, and instructional content of tutor moves. However, the crude metric of the proportion of words used by students vs. teachers does not capture the quality of the conversation; more work can be conducted in this area using the effect estimates from our framework. Together, these patterns point to a practical implication for embedded systems: effectiveness may depend less on maximizing interaction volume and more on ensuring sessions are long enough to resolve a misconception while supporting students in developing productive help-seeking routines. 

Student-level heterogeneity shows larger benefits for learners with lower prior mastery, consistent with just-in-time support models and prior tutoring evidence \cite{nickow2024}. However, low-SES students benefit slightly less on average and show a weaker mastery–impact gradient, aligning with prior work suggesting that on-demand tutoring may require complementary design supports to ensure equitable benefit \cite{kraft2024,deacon2023}.

Finally, the robustness checks show both the value and the limits of using observational data to evaluate embedded tutoring interventions. When we used a broader control group, the estimated effects became smaller, but remained positive. This pattern suggests that the results are not driven by one specific control-group definition. At the same time, it shows how difficult it is to identify the right comparison group when tutoring happens at different moments over time and when earlier tutoring may affect later performance. These challenges underscore the importance of methods that account for students’ changing knowledge states and include diagnostic checks, such as overlap and placebo tests, to assess possible remaining bias \cite{Rubin1974,Glynn2010}.

\section{Limitations \& Future Work}
Several limitations suggest directions for future research. First, although our identification strategy relies on conditional ignorability, sensitivity analyses indicate that unmeasured factors such as student motivation or teacher practices would need to be substantially stronger than our platform-derived measures to overturn the findings. Second, treating tutoring as a binary exposure masks variation in instructional quality and conversational strategy. Future work should use process-level data, such as transcripts or instructional moves, to identify the mechanisms driving effectiveness. Third, our proximal outcomes capture immediate impact and near transfer but not long-term learning gains. Future studies should examine whether session-level effects accumulate over time. Finally, because this study focuses on a single platform, future research should test the framework across domains, platforms, and designs that more directly examine equity and pedagogical mechanisms.
\section{Acknowledgments}
 This research was conducted as part of the National Tutoring Observatory, which was funded by the Bill Gates Foundation (Grant No. INV-068961) and the Chan Zuckerberg Initiative (Grant No. 2024-351541). The views expressed here do not necessarily reflect the official positions or policies of the funding agencies.

%
\bibliographystyle{abbrv}
\balance
\bibliography{sigproc}  
%

\appendix
\href{https://osf.io/vdf8q/files/bez4n?view_only=e45c14c4cc1a4f9e958eeb8e87535ee3}{Digital Appendix}

\end{document}

%% file: 06_tables/sample_flow_table.tex
\begin{table}[t]
\centering
\footnotesize
\setlength{\tabcolsep}{6pt}
\caption{Analytic Sample Construction from Raw Data to Final Analytic Sample}
\label{tab:sample_flow}

\begin{tabular}{@{} >{\raggedright\arraybackslash}p{3.2cm} c c c @{}}
\toprule
 & \textbf{Students} 
 & \shortstack{\textbf{Problem}\\\textbf{Attempts}}
 & \shortstack{\textbf{Unique}\\\textbf{Problems}} \\
\midrule

\textbf{Original Usage Data} & 2,585 & 852,274 & 8,540 \\
\midrule

\multicolumn{4}{@{}l}{\textbf{Treatment Sample}} \\
All Treatment Usage & 1,234 & 465,619 & 6,803 \\
\quad \textit{Excluded} & (13) & (460,456) & (5,337) \\
\textbf{Final Analytic Treated} & 1,221 & 5,163 & 1,466 \\
\midrule

\multicolumn{4}{@{}l}{\textbf{Control Sample}} \\
All Control Usage & 1,351 & 386,655 & 8,017 \\
DKT Training Holdout & 676 & 197,332 & 7,485 \\
Control Analysis & 675 & 189,323 & 5,884 \\
\quad \textit{Excluded} & (10) & (97,874) & (4,418) \\
\textbf{Final Analytic Control} & 665 & 91,449 & 1,466 \\
\midrule

\textbf{Total Analytic Sample} & 1,886 & 96,612 & 1,466 \\
\bottomrule
\end{tabular}

\end{table}

%% file: 06_tables/ate_results_table.tex
\begin{table*}[ht]
\centering
\footnotesize
\caption{Average treatment effect estimates in percentage points across different model specifications.}
\label{tab:ate_results}
\begin{tabular}{lcc}
\toprule
\textbf{Model Specification} & \textbf{ATE (95\% CI)} & \textbf{ATT (95\% CI)} \\
\midrule
\multicolumn{3}{l}{\textit{Primary Specification}} \\
 Immediate Performance (Next Problem) & 4.01\textsuperscript{***} (2.51, 5.51) & 3.98\textsuperscript{***} (2.43, 5.54) \\
 Near Transfer (Next Skill) & 2.73\textsuperscript{**} (1.12, 4.35) & 2.65\textsuperscript{*} (0.99, 4.31) \\
\midrule
\multicolumn{3}{l}{\textit{Robustness Checks}} \\
 External Variables Added (Next Problem) & 3.93\textsuperscript{***} (2.49, 5.37) & 4.71\textsuperscript{***} (2.92, 6.50) \\
 Washout Sample Included  (Next Problem) & 2.17\textsuperscript{***} (1.17, 3.16) & 2.76\textsuperscript{***} (1.91, 3.61) \\
 Placebo Test (Pre-Intervention) & -1.67\textsuperscript{} (-3.10, -0.24) & -0.71\textsuperscript{} (-2.15, 0.74) \\
\bottomrule
\multicolumn{3}{l}{ \textit{Bonferroni-corrected significance:} $^*$ $p < 0.05$, $^{**}$  $p < 0.01$, $^{***}$ $p < 0.001$}
\end{tabular}

\end{table*}